\documentclass{kluwer}
\usepackage{graphics}
\newdisplay{guess}{Conjecture}

\begin{document}
\begin{article}
\begin{opening}

 \title{Discrete Quantum Gravity: II. Simplicial complexes, irreps of SL(2,C), and a Lorentz invariant weight in a state sum model.}

\author{Peter Kramer}
\runningauthor{}
\runningtitle{}
\institute{Institut f\" ur theoretische Physik Universit\" at T\" ubingen, 
72076 T\" ubingen, Germany}
\author{Miguel \surname{Lorente}}
\institute{Departamento de F\'{\i}sica, Universidad de Oviedo, 
 33007 Oviedo, Spain}

\date{April 2, 2008}

\begin{abstract}
In part I of [1] we have developed the tensor and spin representation of SO(4) in order to apply it to the simplicial decomposition of the Barrett-Crane
model. We attach to each face of a triangle the spherical function constructed from the Dolginov-Biedenharn function.

In part II we apply the same technique to the Lorentz invariant state sum model. We need three new ingredients: the classification of the edges and the
corresponding subspaces that arises in the simplicial decomposition, the irreps of SL(2,C) and its isomorphism to the bivectors appearing in the
4-simplices, the need of a zonal spherical function from the intertwining condition of the tensor product for the simple representations attached to the
faces of the simplicial decomposition.
\end{abstract}


\end{opening}    

\section{Introduction.}
Part II deals mainly with the representation theory for relativistic spin networks in quantum gravity.
In section 2 we give some elementary properties of vector and bivector analysis in Minkowski space $M(1,3)$. In section 3 we start from general
representation theory, valid also for the Euclidean version of Part I. Then we describe the global unitary irreps of the group $SL(2,C)$, the universal 
covering group of $SO(1,3,R)$. Section 4 deals in general with 
the notions of intertwiners and invariants, spherical harmonics and
spherical functions. In sections 5, 6, 7 we review the irreps of $SL(2,C)$ and 
obtain the spherical functions occurring in the state sum for relativistic 
spin networks.

\section{Vectors, bivectors, subspaces, and simplices.}

To deal with relativistic spin networks we must go
from the Euclidean space $R^4$ discussed in part I to Minkowski space $M(1,3)$. We sumarize in this section some properties of vectors and bivectors
and use results given in  [3] pp. 56-76.

\noindent{ \bf 1 Def}: A vector $x \in R^n$ is time-like if $ (x\circ x) >0$, space-like
if $ (x\circ x) <0$, light-like if $ (x\circ x) =0$.
A time-like vector is called positive (negative) if $x_1>0\; (x_1<0)$.
We define the vector norm by $|| x||:=\sqrt (x \circ x)$.

\medskip
\noindent {\bf 2 Def}: A simple bivector $b= x \wedge y$ has components
$(x \wedge y)_{\mu \nu}:=x_{\mu}y_{\nu}-x_{\nu}y_{\mu}$
and squared norm
\begin{equation}
\label{e1}
\langle b, b\rangle:= 2((x \circ x)(y \circ y)- (x \circ y)^2).
\end{equation}

\noindent {\bf 3 Def}: [3] p.61: Let  $V$ be  a vector subspace of $R^n$.\\
(1) $V$ is time-like if and only if it has a time-like vector,\\
(2) $V$ is space-like if and only if every nonzero vector of 
$V$ is space-like,\\
(3) $V$ is light-like otherwise.

\medskip
\noindent {\bf 4 Def}: 

(1) $b$ is space-like bivector if $\langle b, b\rangle > 0$, 

(2) $b$ is time-like bivector if $\langle b, b\rangle < 0$,

(3) $b$ is a simple bivector if $\langle b, ^*b\rangle=0$.
\bigskip

\noindent Classification of pairs of vectors by scalar products, span  and 
simple bivectors according to [3]:

\medskip
\noindent {\bf 5.1 Prop}: Let both $x,y$ be positive (negative) time-like. They  span a time-like subspace
$V =\{x,y\}$. Then by [3] p.62 eq. 3.1.6
\begin{eqnarray}
\label{e2} 
&& x \circ y= ||x||\;||y||\; \cosh (x,y),
\\ \nonumber 
&&\langle b, b\rangle = 2 ||x||^2\; ||y||^2 (1-(\cosh(x,y))^2)<0.
\end{eqnarray}
{\bf 5.2 Prop}:  Let $x,y$ be space-like and span a space-like subspace
$V =\{x,y\}$. Then by [3] p.71 eq.  3.2.6
\begin{eqnarray}
\label{e3}
&& x \circ y= ||x||\;||y||\; \cos (x,y),
\\ \nonumber 
&&\langle b, b\rangle = 2 ||x||^2\; ||y||^2 (1-(\cos(x,y))^2)>0.
\end{eqnarray}
{\bf 5.3 Prop}: Let $x,y$ be space-like and span a time-like subspace
$V =\{x,y\}$. Then by [3] p.73 eq. 3.2.7
\begin{eqnarray}
\label{e4}
&& x \circ y= ||x||\;||y||\; \cosh (x,y),
\\ \nonumber 
&&\langle b, b\rangle = 2 ||x||^2\; ||y||^2 (1-(\cosh(x,y))^2)<0.
\end{eqnarray} 
{\bf 5.4 Prop}: Let $x$ be space-like and $y$ be time-like positive.
Then $V =\{x,y\}$ is time-like and by [3] p. 75 eq.  3.2.8
\begin{eqnarray}
\label{e5}
&& x \circ y = ||x||\;||y||\; \sinh (x,y),
\\ \nonumber 
&&\langle b, b\rangle = 2 ||x||^2\; |||y|||^2 (1-(\sinh(x,y))^2)={\rm indefinite}.
\end{eqnarray}

\noindent {\bf 6 Def}: An m-simplex $S^m$ with a vertex at $0$ can be defined as  
the convex hull of the points from
a set of $m$ linearly independent edge vectors $x^1,x^2,\ldots, x^m$ as
$S^m := {\rm conv} (\lambda_1 x^1, \lambda_2 x^2,\ldots,\lambda_m x^m),
0 \leq \lambda_j\leq 1$.
Then we can generalize  Prop 1.2 to

\medskip
\noindent 
{\bf 7 Prop}: Let the m-simplex $S^m, m < n$ be the convex hull of 
$m$ space-like linearly independent 
edge vectors $x^1,x^2,\ldots, x^m$ which span a space-like subspace $V$.
Then all subsimplices $S^p, p<m$ of $S^m$ have space-like edge vectors  
which span space-like subspaces.

{\em Proof}: All the edge vectors of the simplex $S^m$ are in the space-like 
subspace spanned by $x^1,x^2,\ldots, x^m$. Any subspace of a space-like 
subspace is space-like. We can prove the existence of these simplices 
by choosing in Minkowski space a set of up to $n-1$ orthogonal space-like
unit vectors. Certainly these span a space-like subspace.

For the case $m=n$ we can still choose $n$ space-like linearly independent 
vectors. But as these provide a basis for the full Minkowski space, they 
must include time-like vectors and so cannot span a space-like subspace.

\medskip
\noindent 
{\bf 8 Prop}: Classification of simple bivectors $b= a_1 \wedge a_2$  
attached to a triangle  by their scalar product:

\medskip
\noindent 
If $a_1,a_2$ span a space-like subspace then from {\bf Prop 5.2}: $\langle b, b \rangle >0$. 

\medskip
\noindent 
If $a_1,a_2$ span a time-like subspace then from {\bf Prop 5.1 and 5.3}: \mbox{$\langle b, b \rangle <0$}.

\medskip
\noindent 
{\bf 9 Prop}: The second implication cannot be reversed.
Only if we exclude simple bivectors formed from a space-like and a time-like
vector then  the squared norm of the bivector determines the type of
spanned subspace.

We see from these results that, for a simplicial triangulation in Minkowski space $M(1,3)$, care must be taken already with the choice of triangular faces and bivectors. We are not aware of considerations of this point 
in the literature on relativistic spin networks. 

\section{Irreducible unitary representations of $SL(2,C)$.}

When we switch from Euclidean space $R^4$ to Minkowski space $M(1,3)$,
we must deal with the unitary irreps of the group $SL(2,C)$ which is
the universal covering group of the group $SO(1,3,R)$. We start 
in subsection 3.1 with some general properties of representations and then describe in subsections 3.2 and 3.3 the unitary irreps of $SL(2,C)$.

\subsection{Operators, automorphisms and irreducible representations.}
Consider in a Hilbert space of complex-valued functions of a complex 
variable with scalar product
\begin{equation}
\label{e6}
\langle \phi, \psi \rangle = \int \overline{\phi(z)} \psi(z) d\mu(z),\;
d\mu(z):=dRe(z)dIm(z)
\end{equation}
a set $\{ T \}$ of linear operators $T$ with inverses $T^{-1}$. 
Denote complex conjugation of functions by overlining. 

\medskip
\noindent 
{\bf 10 Def}: Define the transposed, the conjugate
and the adjoint operator respectively by
\begin{eqnarray}
\label{e7}
&T^T:&\;  \langle \phi, T^T \psi \rangle= \langle \psi, T \phi\rangle,
\\ \nonumber
&T^C:&\; \overline{\langle  \phi, T \psi\rangle} 
=\langle \phi', T^C \psi' \rangle,
\\ \nonumber
&T^{\dagger}:&\;:\langle \phi, T^{\dagger} \psi \rangle
= \langle T \phi, \psi\rangle.
\end{eqnarray}
The states $\psi',\phi'$ are defined  in eq. \ref{e23} below.

The following involutive operator automorphims $\Sigma: T \rightarrow \Sigma(T)$ 
obeying $\Sigma (T_{\alpha}T_{\beta})= \Sigma (T_{\alpha})\Sigma (T_{\beta})$
are called [2] p.139 contragredient, conjugate and 
contragredient-conjugate:
\begin{eqnarray}
\label{e8}
&& \Sigma_1:\; T \rightarrow (T^T)^{-1}=(T^{-1})^T, 
\\ \nonumber
&& \Sigma_2:\; T \rightarrow T^C, 
\\ \nonumber
&& \Sigma_3:\; T \rightarrow (T^{\dagger})^{-1}=(T^{-1})^{\dagger}. 
\end{eqnarray}
Next consider a representation of a group $G$ by the set  $\{ T \}$ 
of operators,
\begin{equation}
\label{e9}
g \in G \rightarrow T_g,\: T_e=I, \;T_{g^{-1}}=(T_g)^{-1},\; T_{g_1g_2}= T_{g_1}T_{g_2}.
\end{equation}
If the representation is unitary we find
\begin{equation}
\label{e10}
(T^{\dagger}_g)^{-1}=T_g,\: T_g^C=(T^{-1}_g)^T, \; \Sigma_3(T)=T,
\: \Sigma_1(T)=\Sigma_2(T).
\end{equation}
Denote by $g \in G \rightarrow T^{\lambda}_g$ a  unitary irreducible representation
with irrep label $\lambda$. By the operator automorphisms eq. \ref{e8}
and by unitarity eq. \ref{e10} it follows that for any irrep $\lambda$ there
is a conjugate irrep $(T^{\lambda}_g)^C $  whose irrep label we denote by $\lambda^C$. 

\subsection{Principal series of irreducible representations of $SL(2,C)$.} 

Consider now a group of complex-valued matrices $g \in G$. With respect to
this group we have in analogy to eq.\ref{e8} three involutive 
group automorphisms
obeying $\sigma: \sigma(g_1g_2)=\sigma(g_1)\sigma(g_2)$,
\begin{eqnarray}
\label{e11}
&\sigma_1:&\; g \rightarrow (g^{-1})^T,
\\ \nonumber
&\sigma_2:&\; g \rightarrow \overline{g},
\\ \nonumber
&\sigma_3:&\; g \rightarrow (g^{-1})^{\dagger}.
\end{eqnarray}
The three group automorphisms eq. \ref{e11} allow to construct from
a fixed representation $ g \rightarrow T_g$ three new representations
$T_{\sigma_1(g)}$, $T_{\sigma_2(g)}$,
$T_{\sigma_3(g)}$. 

These representations should be carefully distinguished from the ones
defined by eq. \ref{e8}, since the automorphisms  of operators eq. \ref{e7} in representation space 
in general cannot be 
pulled down to automorphisms eq. \ref{e11} of  group elements, 
as will be seen in eq. \ref{e29} below! 

We define the complex valued polynomials
$$p(z,\bar z) = \sum_{}^{} C_{\alpha \beta }{ z^\alpha}  {\bar z^\beta }$$
as the basic states of the spinor representations:

$$T_a p(z,\bar z) = \left( {\beta z + \delta } \right)^k \left( {\bar \beta \bar z + \bar \delta } \right)^n p\left( {{{\alpha z + \gamma } \over {\beta z + \delta
}},{{\bar \alpha \bar z + \bar \gamma } \over {\bar \beta \bar z + \bar \delta }}} \right)$$
for any $$a \equiv \left(
\begin{array}{ll}
\alpha & \beta \\
\gamma & \delta
\end{array}
\right) \in SL(2,C),\alpha \delta  - \beta \gamma  = 1$$.

This representation with labels $\left( {l_0 ,l_1 } \right) = \left( {{{k - n} \over 2},{{k + n} \over 2} + 1} \right),\; \; k,n \in N$, is irreducible and finite
dimensional, see [1] eq. (20). If we enlarge this representation with complex values, $k,n$ we get:

$$T_a f(z,\bar z) = \left( {\beta z + \delta } \right)^{l_0  + l_1  - 1} \left( {\bar \beta \bar z + \bar \delta } \right)^{l_0  - l_1  - 1} f\left( {{{\alpha z +
\gamma } \over {\beta z + \delta }},{{\bar \alpha \bar z + \bar \gamma } \over {\bar \beta \bar z + \bar \delta }}} \right)$$
the representation becomes infinite dimensional. For $l_0$ integer or half integer and $l_1$ arbitrary complex, the representation is irreducible.

We follow [2] p. 565 and define the principal series of irreps of
$SL(2,C)$ on a Hilbert space of functions in a complex variable 
$z$ with scalar product eq. \ref{e6}.

\medskip
\noindent{\bf 11 Def}: For the group element
\begin{equation}
\label{e12}
g=
\left[
\begin{array}{ll}
\alpha & \beta \\
\gamma & \delta
\end{array}
\right],\: g^{-1}=
\left[
\begin{array}{ll}
\delta & -\beta \\
-\gamma & \alpha
\end{array}
\right]
\: \in SL(2,C),\; \alpha \delta-\beta\gamma=1,
\end{equation}
we adopt the operator 
\begin{equation}
\label{e13}
(T^{\left[m, \rho \right]}_{g}\psi)(z)
:=|\beta z+ \delta|^{m+i\rho-2}(\beta z+\delta)^{-m}
\psi\left( \frac{\alpha z +\gamma}{\beta z + \delta}\right).
\end{equation}
The representation is characterized by the integer/real numbers 
$\left[ m,\rho\right]$. These numbers determine the eigenvalues of
two Casimir operators $C_1, C_2$,  [4] p.167 according
to
\begin{equation}
\label{e14}
C_1\;\psi^{\left[m,\rho\right]}=
-\frac{(m^2-\rho^2-4)}{2}\psi^{\left[m,\rho\right]},
C_2\;\psi^{\left[m,\rho\right]}=
m\rho\psi^{\left[m,\rho\right]}.
\end{equation}
For the inverse element we find
\begin{equation}
\label{e15}
(T^{\left[m, \rho \right]}_{g^{-1}}\psi)(z)
=|-\beta z+ \alpha|^{m+i\rho-2}(-\beta z+\alpha)^{-m}
\psi\left( \frac{\delta z -\gamma}{-\beta z + \alpha}\right) 
\end{equation}
To find the adjoint operator we substitute in the integral expression for
the left-hand side of 
\begin{equation}
\label{e16}
\langle \phi, T_g^{\left[m,\rho\right]}\psi \rangle 
= \langle (T^{\left[m,\rho\right]}_g)^{\dagger}\phi, \psi \rangle
\end{equation}
the new variable 
\begin{equation}
\label{e17}
z'=\frac{\alpha z +\gamma}{\beta z + \delta}.
\end{equation}
To transform the measure $d\mu(z)$ we compute
\begin{eqnarray}
\label{e18}
&&\frac{\partial z'}{\partial z}=1/\;(\beta z + \delta)^2,
\\ \nonumber 
&&\frac{\partial \overline{z}'}{\partial \overline{z}}
=1/\;\overline{(\beta z + \delta)^2}, 
\\ \nonumber 
&&\frac{d\mu(z')}{d\mu(z)}
=1/\;|\;\beta z+\delta|^4 = |-\beta z' +\alpha|^4.
\end{eqnarray}
After some rewriting we obtain for the adjoint operator 
\begin{equation}
\label{e19}
(T^{\left[m,\rho\right], \dagger}_g\phi)(z')
= |-\beta z'+ \alpha|^{m+i \rho-2}(-\beta z'+\alpha)^{-m}
\phi\left( \frac{\delta z' -\gamma}{-\beta z' + \alpha}\right) .
\end{equation} 
Comparing with the operator $T_{g^{-1}}$ (15) we verify the 
unitarity eq. \ref{e10} of the representation eq. \ref{e13},
\begin{equation}
\label{e20}
(T^{\left[m,\rho \right]}_g)^{\dagger}
= T^{\left[m,\rho \right]}_{g^{-1}}.
\end{equation}
From unitarity it follows by eq. \ref{e10}, that, besides of the
original irrep eq. \ref{e13}, we need to consider only 
the conjugate irrep. 
To determine the operation of conjugation it is better to consider 
the elements  $\psi, \phi$ as functions of $z, \overline{z}$.
Then eq. \ref{e13} becomes
\begin{equation}
\label{e21}
(T^{\left[m, \rho \right]}_{g}\psi)(z,\overline{z})
:=|\beta z+ \delta|^{m+i\rho-2}(\beta z+\delta)^{-m}
\psi(\frac{\alpha z +\gamma}{\beta z + \delta},
\frac{\overline{\alpha}\overline{z} +\overline{\gamma}}
{\overline{\beta} \overline{z} + \overline{\delta}}).
\end{equation}
The complex conjugation of this equation is
\begin{equation}
\label{e22}
\overline{(T^{\left[m, \rho \right]}_{g}\psi)(z,\overline{z})}
:=|\overline{\beta} \overline{z}+ \overline{\delta}
|^{m-i\rho-2}(\overline{\beta} 
\overline{z}+\overline{\delta})^{-m}
\overline{\psi(\frac{\alpha z +\gamma}{\beta z + \delta},
\frac{\overline{\alpha}\overline{z} +\overline{\gamma}}
{\overline{\beta} \overline{z} + \overline{\delta}})}.
\end{equation}
We wish to rewrite this expression as the action of a conjugate 
operator on a complex-valued function. To this purpose we define
\begin{equation}
\label{e23} 
\overline{\psi(z,\overline{z})}=: \psi'(z',\overline{z}'),\;
\overline{\phi(z,\overline{z})}=: \phi'(z',\overline{z}'),\;
z'=\overline{z}, \overline{z}'= z.
\end{equation}
These definitions allow to rewrite eq. \ref{e22} as
\begin{eqnarray}
\label{e24}
&&\overline{(T^{\left[m, \rho \right]}_{g}\psi)(z,\overline{z})}
\\ \nonumber
&&:=|\overline{\beta} z'+ \overline{\delta}|^{m-i\rho-2}
(\overline{\beta} z'+\overline{\delta})^{-m}
\psi'(\frac{\overline{\alpha} z' +\overline{\gamma}}
{\overline{\beta} z' + \overline{\delta}},
\frac{\alpha\overline{z}' +\gamma}
{\beta \overline{z}' + \delta}).
\end{eqnarray}
Comparing with the definition eq. \ref{e21} we determine 
the conjugate operator as
\begin{equation}
\label{e25}
\overline{(T^{\left[m, \rho \right]}_{g}\psi)(z,\overline{z})}
:=(T^{\left[m, \rho \right],C}_{g}\psi')(z',\overline{z}')
=(T^{\left[m, -\rho \right]}_{\sigma_2(g)}\psi')(z',\overline{z}'),
\end{equation}
and obtain for the conjugate irrep 
$\left[m,\rho\right]^C=\left[m,-\rho\right]$.
Moreover  with eq. \ref{e10} one easily verifies
\begin{equation}
\label{e26}
T^{\left[m, \rho \right],C}_g
=:(T^{\left[m, \rho \right]}_g)^C
=(T^{\left[m, \rho \right]}_{g^{-1}})^T 
= T^{\left[m, -\rho \right]}_{\sigma_2(g)}.  
\end{equation}
We keep the group automorphism $\sigma_2$ from eq. \ref{e11}
since we wish to consider the  expressions in eq. \ref{e26}  
as homomorphisms $g \rightarrow T_g$.

\medskip
\noindent{\bf 12 Prop}: The irreps labelled by $\left[m,\rho\right]$ eq. \ref{e13} of the principal series of $SL(2,C)$
are unitary. The conjugate irreps are labelled by $\left[m,-\rho\right]$
and given by eq. \ref{e25}.


Finally we turn to the group automorphisms eq. \ref{e11}. 
For the group $SL(2,C)$ we have the particular result 
\begin{equation}
\label{e27}
(g^{-1})^T= q g q^{-1},\; 
q=-q^{-1}=\overline{q}= 
\left[
\begin{array}{ll}
0 &1 \\
-1& 0
\end{array}
\right].
\end{equation}
Since $q \in SL(2,C)$, $\sigma_1$ becomes an inner automorphism.
eq. \ref{e11} implies
\begin{equation}
\label{e28}
\sigma_1(g)=q\; g\: q^{-1},\; 
\sigma_3(g)=\overline{q}\; \overline{g}\; \overline{q^{-1}}
= q\; \overline{g}\; q^{-1}= q\; \sigma_2(g)\; q^{-1}   
\end{equation}
The automorphisms $\sigma_2, \sigma_3$ are related by conjugation
in the group.
In terms of the irreducible representations, we find from these 
expressions 
\begin{eqnarray}
\label{e29}
&&T^{\left[m,\rho\right]}_{(g^{-1})^T}= 
T^{\left[m, \rho \right]}_q\; 
T^{\left[m, \rho \right]}_g\;
T^{\left[m, \rho \right]}_{q^{-1}}\; \sim T^{\left[m, \rho \right]}_g,
\\ \nonumber
&&T^{\left[m, \rho \right]}_{\overline{g}}
= (T^{\left[m, -\rho \right]}_{g})^C,
\\ \nonumber
&&T^{\left[m, \rho \right]}_{(g^{-1})^{\dagger}}
=T^{\left[m, \rho \right]}_q\; 
T^{\left[m, \rho \right]}_{\overline{g}}\;
T^{\left[m, \rho \right]}_{q^{-1}} 
\sim T^{\left[m, \rho \right]}_{\overline {g}}.
\end{eqnarray}

\subsection{Complementary series of irreducible representations of SL(2,C).}
We consider the Hilbert space of complex-valued functions of a complex variable with scalar product
\begin{equation}
\label{e30}
\left( {f_1 ,f_2 } \right) = \int\!\!\!\int {\left| {z_1  - z_2 } \right|} ^{ - 2 + \sigma } f_1 (z_1 )\overline {f_2 (z_2 )} dz_1 dz_2 
\end{equation}
where the transposed, conjugate and adjoint operators are defined as before in eq. \ref{e7}, and similarly the contragradient, the conjugate and the contragradient conjugate
involutive automorphisms are defined as in eq. \ref{e8}.

Following [2] page 573, we define the complementary series of irreps of SL(2,C) on a Hilbert space with scalar product eq. \ref{e30}.

\medskip
\noindent
{\bf 13 Def}: For the group element
\begin{equation}
\label{e31}
g = \left(
\begin{array}{cc}
\alpha & \beta \\
\gamma & \delta
\end{array}
\right) , g^{-1}= 
\left(
\begin{array}{cc}
\delta & -\beta \\
-\gamma & \alpha
\end{array}
\right)\in SL(2,C),\alpha \delta  - \beta \gamma  = 1,
\end{equation}
we adopt the operator expression
\begin{equation}
\label{e32}
\left( {T_g^{[0,\sigma ]} \psi } \right)(z) = \left| { \beta z + \delta } \right|^{ - 2 - \sigma } \psi \left( {{{\alpha z + \gamma } \over { \beta z + \delta
}}} \right)
\end{equation}
and the inverse operator
\begin{equation}
\label{e33}
\left( {T_{g^{ - 1} }^{[0,\sigma ]} \psi } \right)(z) = \left| { - \beta z + \alpha } \right|^{ - 2 - \sigma } \psi \left( {{{\delta z - \gamma } \over { - \beta
z + \alpha }}} \right),
\end{equation}
where $\sigma$ is a real number $0<\sigma<2.$

The eigenvalues of the two Casimir operators are
\begin{equation}
\label{e34}
C_1 \psi ^{[0,\sigma ]}  = -{{\sigma ^2  - 4} \over 2}\psi ^{[0,\sigma ]} , \; \; \; C_2 \psi ^{[0,\sigma ]}  = 0.
\end{equation}
To find the adjoint operator, we substitute in  the integral expression eq. \ref{e7} with the help of eq. \ref{e30} the new variables
\begin{equation}
\label{e35}
z'_1  = {{\alpha z_1  + \gamma } \over {\beta z_1  + \delta }}\; \; ,\; \; z'_2  = {{\alpha z_2  + \gamma } \over {\beta z_2  + \delta }}
\end{equation}
with the transformed measure
\begin{equation}
\label{e36}
dz_1  = {{dz'_1 } \over {\left( { - \beta z'_1  + \alpha } \right)^4 }}\; \; ,\; \; dz_2  = {{dz'_2 } \over {\left( { - \beta z'_2  + \alpha } \right)^4 }}.
\end{equation}
After simplification we obtain
\begin{equation}
\label{e37}
\left( {T_{g^{} }^{[0,\sigma ]\dagger} \psi } \right)(z') = \left| { - \beta z' + \alpha } \right|^{ - 2 - \sigma } \psi \left( {{{\delta z' - \gamma } \over { -
\beta z' + \alpha }}} \right).
\end{equation}
Comparing with eq. \ref{e33} we verify the unitarity of the representation eq. \ref{e32}
\begin{equation}
\label{e38}
T_{g^{} }^{[0,\sigma ]\dagger}  = T_{g^{ - 1} }^{[0,\sigma ]} 
\end{equation}
To determine the operation of conjugation we follow the same steps eq. \ref{e21} to 
eq. \ref{e26} for the principal series, keeping in mind that $m=0, \rho=i \sigma$ is a a pure
imaginary number. We obtain
\begin{equation}
\label{e39}
T_{g^{} }^{[0,\sigma ]C}  = T_{\sigma _2 (g)^{} }^{[0,\sigma ]},
\end{equation}
where $\sigma_2$ is defined in eq. \ref{e11}.

\section{Right action invariants from two irreps.}

In this section we use representation theory to 
examine  the notions of intertwining and invariants as they appear in
relation with discrete quantum gravity.

We switch for convenience to a bracket notation. Our results hold true 
for general unitary irreps of a group and in particular for the
irreps of $SO(4,R)$ analyzed in part I.
Consider for the unitary irrep $\lambda =\left[ m,\rho\right]$ and two group elements
$g_1, g_2$ the matrix element 
\begin{equation}
\label{e40}
\langle \lambda \mu| T^{\left[m, \rho \right]}_{g_1g_2^{-1}}|
\lambda \mu'\rangle
= \sum_{\nu}
\langle \lambda \mu|T^{\left[m, \rho \right]}_{g_1}|\lambda \nu\rangle
\langle \lambda \nu|T^{\left[m, \rho \right]}_{g_2^{-1}}|\lambda \mu'\rangle.
\end{equation}
By construction this matrix element is invariant under the right action
\begin{equation}
 \label{e41}
g_1 \rightarrow g_1g,\; g_2 \rightarrow g_2 g.
\end{equation}  
We transform the matrix elements for the group element $g_2$ 
by 
\begin{eqnarray}
\label{e42}
&&\langle \lambda \nu|T^{\lambda}_{g_2^{-1}}|\lambda \mu'\rangle
=\langle \lambda \nu|(T^{\lambda}_{g_2})^{\dagger}|\lambda \mu'\rangle
\\ \nonumber
&&= \overline{\langle \lambda \mu'|T^{\lambda}_{g_2}|\lambda \nu\rangle}
= \langle \lambda^C {\mu'}^C|
T^{\lambda^C}_{\sigma_2(g_2)}
|\lambda^C \nu^C\rangle.
\end{eqnarray}
In the last line we introduced the conjugate irrep and the 
superscripts $C$ for its row and column labels. Substituting eq. \ref{e42} into 
eq. \ref{e40}
and using eq. \ref{e26} we get
\begin{eqnarray}
\label{e43}
&&\langle \lambda \mu| T^{\lambda}_{g_1g_2^{-1}}|
\lambda \mu'\rangle
\\ \nonumber
&& = \sum_{\nu}\langle \lambda \mu|T^{\lambda}_{g_1}|\lambda \nu\rangle
\langle \lambda^C \mu'^C|
T^{\lambda^C}_{\sigma_2(g_2)}
|\lambda^C \nu^C\rangle.
\end{eqnarray}
Next we wish to  pass from the invariant eq. \ref{e40} to
the intertwining or Kronecker coupling of two conjugate irreps.
Consider a right-hand Kronecker coupling of two irreps $\lambda, \lambda'$
for  two different group elements
$g_1, g_2$, and its decomposition 
under the right actions eq. \ref{e41},
\begin{eqnarray}
\label{e44}
\nonumber &&\sum_{\nu, \nu'}
\langle \lambda \mu| T^{\lambda}_{g_1g}|\lambda \nu\rangle
\langle \lambda' \mu'| T^{\lambda'}_{g_2g}|\lambda' \nu'\rangle
\langle \lambda \nu, \lambda' \nu'| \lambda^* \mu^*
\rangle 
\\ 
&=&\sum_{\sigma,\sigma'} \sum_{\nu,\nu'}
\langle \lambda \mu| T^{\lambda}_{g_1}|\lambda \sigma\rangle
\langle \lambda' \mu'| T^{\lambda'}_{g_2}|\lambda' \sigma'\rangle
\langle \lambda \sigma| T^{\lambda}_{g}|\lambda \nu\rangle
\\ \nonumber
&&\langle \lambda' \sigma'| T^{\lambda'}_{g}|\lambda' \nu'\rangle
\langle \lambda \nu, \lambda' \nu'| \lambda^* \mu^*\rangle.
\end{eqnarray}
In this expression the sum over Wigner coefficients and  functions of $g$ only yields
\begin{eqnarray}
\label{e45}
&&\sum_{\nu,\nu'}
\langle \lambda \sigma| T^{\lambda}_{g}|\lambda \nu\rangle
\langle \lambda' \sigma'| T^{\lambda'}_{g}|\lambda' \nu'\rangle
\langle \lambda \nu, \lambda' \nu'| \lambda^* \mu^*\rangle
\\ \nonumber
&&= \sum_{\mu^{*'}}
\langle  \lambda \sigma, \lambda' \sigma' |\lambda^* \mu^{*'}\rangle
\langle \lambda^* \mu^{*'} |T^{\lambda^*}_g| \lambda^* \mu^* \rangle.
\end{eqnarray} 
The only term on the right-hand side of this expression independent of $g$ 
is the one with 
$(\lambda^*,\mu^*)=(\lambda^*,\mu^{*'})=(0,0)$ whose Wigner coefficients couple to the
identity irrep $\lambda=0$. Therefore if we make this choice in the left hand side of the
first line in eq. \ref{e44} we must get an expression invariant under the
substitution eq. \ref{e41},
\begin{equation}
\label{e46}
\sum_{\nu, \nu'}
\langle \lambda \mu| T^{\lambda}_{g_1}|\lambda \nu\rangle
\langle \lambda' \mu'| T^{\lambda'}_{g_2}|\lambda' \nu'\rangle
\langle \lambda \nu, \lambda' \nu'| 00\rangle.
\end{equation}
By its form and invariance, this expression must coincide with eq. \ref{e40}
up to a constant.
Comparing these expressions we find the Wigner coefficients 
for eq. \ref{e46} up to normalization: 

\medskip
\noindent
{\bf 14 Prop}:  The Kronecker coupling to an invariant requires two conjugate 
representations  $\lambda, \lambda^C$ of eq. \ref{e46}
and has the Wigner coefficients
\begin{eqnarray}
 \label{e47}
&&\lambda'= \lambda^C,\; 
\left[ m', \rho'\right]= \left[ m,-\rho\right],
\\ \nonumber
&& \langle \lambda \nu, \lambda' \nu'| 00\rangle
= \delta(\lambda',\lambda^C) \delta(\nu',\nu^C).
\end{eqnarray}
The simplicity of this expressions results from the introduction of
the conjugate representation labels in eq. \ref{e42}. For a general analysis 
of Wigner coefficients for $SL(2,C)$ principal series representations
we refer to [6].

\section{Representations of the algebra of SL(2,$\mathbb{C}$)}

In order to discuss relativistic spherical harmonics and zonal spherical functions we must
write the irreps of $SL(2,C)$ as given in eqs. \ref{e12}, \ref{e13} in a form which is explicitly reduced under the subgroup
$SU(2)$.  This will give the relativistic counterpart of the Gelfand-Zetlin basis 
used in the Euclidean case of part I.

For this purpose   we expand the basis $\psi ^{\left[ {m,\rho } \right]}$ in terms of the orthogonal Hahn
polynomials of imaginary argument [5] [7] with the help of complexified Clebsch-Gordan coefficients.

Given the generators  $\left( {J_1,J_2,J_3} \right)=\vec J$ of $SU(2)$ and of pure Lorentz transformations $\left(
{K_1,K_2,K_3} \right)=\vec K$ satisfying the commutation relations:
\begin{equation}
\label{e52}
\begin{array}{llllll}
&\left[ {J_p,J_q} \right]=i\varepsilon_{pqr}\,J_r\quad &,\quad &\vec J^+=\vec J\quad &,\quad &p,q,r=1,2,3\cr
  &\left[ {J_p,K_q} \right]=i\varepsilon_{pqr}\,K_r\quad &,\quad &\vec K^+=\vec K\cr
  &\left[ {K_p,K_q} \right]=-i\varepsilon_{pqr}\,J_r\cr
\end{array}
\end{equation}
we obtain the unitary representations of the algebra of $SL(2, \mathbb{C})$ in the basis where the operators $J_3$ and $\vec J^2$ are diagonal,
namely: $J_3\psi _{JM}=M\psi _{JM},\quad \vec J^2\psi _{JM}=J(J+1)\psi _{JM}$

It is possible also to construct complexified operators
$$\vec A={1 \over 2}\left( {\vec J+i\vec K} \right),\quad \vec B={1 \over 2}\left( {\vec J-i\vec K} \right),\quad \vec A^+=\vec B$$
that leads to the commutation relations of two independent angular momenta:
\vspace{-0,4 cm}
\begin{equation}
\label{e53}
\begin{array}{l}
\left[ {A_p,A_q} \right]=i\varepsilon_{pqr}\,A_r \\
\left[ {B_p,B_q} \right]=i\varepsilon_{pqr}\,B_r \\
\left[ {A_p,B_q} \right]=0
\end{array}
\end{equation}

Since $J_3$ and $K_3$ commute we construct for a fixed irrep $[m,\rho]$ the representations of these operators in the basis where $J_3$ and $K_3$ are diagonal [5],

\vspace{0,1 cm}
$J_3\phi _{m_1m_2}=M\phi _{m_1m_2}\quad ,\quad K_3\phi _{m_1m_2}=\lambda \phi _{m_1m_2},$ hence

$A_3\phi _{m_1m_2}={1 \over 2}\left( {M+i\lambda } \right)\phi _{m_1m_2}\equiv m_1\phi _{m_1m_2}$

$B_3\phi _{m_1m_2}={1 \over 2}\left( {M-i\lambda } \right)\phi _{m_1m_2}\equiv m_2\phi _{m_1m_2}$
\vspace{0,2 cm}

Notice that $\lambda$ is a real continuous parameter, but $m_1$ and $m_2$ are complex conjugate and $\bar m_1=m_2$

For the Casimir operators we have
\vspace{-0,4 cm}
$$
\begin{array}{l}
C_1=\left( {\vec J^2- \vec K^2} \right)\psi _{JM}=\left( {l_0^2+l_1^2-1} \right)\psi _{JM} \\
C_2=\left( {\vec J\cdot \vec K} \right)\psi _{JM}=l_0l_1\psi _{JM} 
\end{array}
$$
or in the $\left[ {m,\rho } \right]$ notation
\begin{equation}
\label{e54}
\begin{array}{l}
C_1 \psi ^{\left[ {m,\rho } \right]}  = {1 \over 2}\left( {m^2  - \rho ^2  - 4} \right)\psi ^{\left[ {m,\rho } \right]} \\
C_2 \psi ^{\left[ {m,\rho } \right]}  = m\rho \psi ^{\left[ {m,\rho } \right]}
\end{array}
\end{equation}
We shall extend the analysis to the irreps $[0,\sigma]$ of the completely degenerate series 
eq. \ref{e32}.

\section{Complexified Clebsch-Gordan coefficients and the representation of the boost operator}

In order to connect the basis $\psi _{JM}$ and $\phi _{m_1m_2}$ we can use the complexified Clebsch-Gordan coefficients:
\begin{equation}
\label{e55}
\psi _{JM}=\int\limits_{-\infty }^\infty  {d\lambda \left\langle {{m_1m_2}} \mathrel{\left | {\vphantom {{m_1m_2} {JM}}} \right.
\kern-\nulldelimiterspace} {{JM}} \right\rangle }\phi _{m_1m_2}
\end{equation}
We have used integration because $\lambda$  is a continuous parameter. It can be proved that these coefficientes are related to the Hahn
polynomials of imaginary argument [5]
\begin{equation}
\label{e56}
\left\langle {{m_1m_2}} \mathrel{\left | {\vphantom {{m_1m_2} {j_m}}} \right. \kern-\nulldelimiterspace} {{JM}} \right\rangle =f{{\sqrt {\omega
(\lambda )}} \over {d_{J-M}}}p_{J-M}^{(M-m ,M+m )}(\lambda ,\rho ),\: m=m_1+m_2, 
\quad f \overline{f}=1,
\end{equation}
for the principal series,
\begin{equation}
\label{e56a}
\left\langle {{m_1m_2}} \mathrel{\left | {\vphantom {{m_1m_2} {j_m}}} \right. \kern-\nulldelimiterspace} {{JM}} \right\rangle =f\sqrt {\omega
(\lambda )}\;d_{J-M}^{-1}\;q_{J-M}^{(M)}(\lambda ,\sigma ),\: m=m_1+m_2, 
\quad f\overline{f}=1
\end{equation}
for the complementary series, 

\noindent where
\begin{equation}
\label{e57}
\omega (\lambda) ={1 \over {4\pi }}\left| \Gamma \left( {{M-m +1} \over {2}} +{i{\lambda -\rho } \over {2} }\right)\Gamma \left( {{m+\mu +1} \over {2}}+i{{\lambda +\rho } \over {2}}\right)\right|^2
\end{equation}
\noindent and
\begin{equation}
\label{e58}
(d_n)^2={{\Gamma \left( {M-m +n+1} \right)\Gamma \left( {M+m +n+1} \right)\left| {\Gamma \left( {M+i\rho +n+1}
\right)} \right|^2} \over {n!\;\left( {2M +2n+1} \right)\Gamma \left( {2M +n+1} \right)}}
\end{equation}

\noindent for the principal series and similar expression for the complementary series [5].

With the help of  eqs. \ref{e55}, \ref{e56} and \ref{e56a} we can construct the
representation for the boost operator, or the Biedenharn-Dolginov function, namely [5],

\begin{eqnarray}
\label{e59}
&&d^{[m,\rho]}_{JJ'M}(\tau)\\
\nonumber 
&&=\int_{-\infty}^{\infty} d_{J-M}^{-1} p_{J-M}^{(M-m,M+m)}(\lambda,\rho)
\exp(-i\tau\lambda)d_{J-M}^{-1} p_{J'-M}^{(M-m,M+m)}(\lambda,\rho)
\omega(\lambda)d\lambda
\\ \nonumber 
&&{\rm for}\: {\rm the}\: {\rm principal}\:  {\rm series},\\  
\nonumber 
&&d^{[0,\sigma]}_{JJ'M }(\tau) 
\\ \nonumber
&&= \int_{ - \infty }^\infty  d_{J-M}^{ - 1} q_{J-M}^{(M)} (\lambda ,\sigma )
\exp(- i\tau \lambda) d_{J' - M}^{-1} q_{J' - M}^{(M)}
(\lambda ,\sigma )\omega (\lambda )d\lambda 
\\ \nonumber 
 &&{\rm for}\: {\rm the}\: {\rm complementary}\: {\rm series}.
\end{eqnarray}

\section{Relativistic spherical harmonics on $SL(2,C)/SU(2)$, intertwining, and spherical functions.}

We consider irreps and spherical harmonics on the homogeneous space 
$SL(2,C)/SU(2)$. Here $SU(2)$ is the stability group of the point $P_0=(1,0,0,0) \in M(1,3)$,
and the homogeneous space is the hyperboloid $H^3 < M(1,3)$ which replaces the 3-sphere 
$S^3< R^4$ used in part I [1].
We show that the general intertwining of pairs analyzed in
section 4  from  spherical functions yields zonal spherical functions.

Consider those  irreps $\lambda$ of $SL(2,C)$ which admits once and only 
once the irrep $J=0$ of the compact subgroup $SU(2)$. An orthogonal
irrep set of functions on the coset space $SL(2,C)/SU(2)$ can be taken 
as the set of matrix elements 
\begin{equation}
\label{e48}
\langle [m,\rho]  (JM=00) | T^{\lambda}_g| [m, \rho] (J'M')\rangle
=: \langle [m,\rho] \downarrow(00)| T^{\lambda}_g| [m,\rho] (J'M')\rangle,
\end{equation}
where we introduced $\downarrow(00)$ for the subduction to the identity  irreps
of the subgroup $SU(2)$ to distinguish it from the notation $(00)$ used 
for the identity irreps of the full group $SL(2,C)$ in eq. 
\ref{e46}. Clearly the matrix elements 
in eq. \ref{e48} are invariant under the substitution 
$g \rightarrow ug,\; u \in SU(2)$ and so may be assigned to the left cosets 
$SL(2,C)/SU(2)$. One could decompose the measure on
group space and from the orthogonality of
the irreps obtain an orthogonality rule for these irrep functions
which play the role of relativistic spherical harmonics. These relativistic spherical harmonics for 
$SO(1,3,R)$ or
$SL(2,C)$ must be distinguished from the ones of the group $SO(4,R)$
discussed in part I section 5.2 and Theorem 1. Vilenkin and Klimyk in [10]  vol. 2 p. 28
discuss the relativistic spherical harmonics  in relation with $\Delta$- and $\Box$-harmonics.

In the models of discrete relativistic quantum gravity, these spherical harmonics  can be assigned
as fields to the 3-simplices or tetrahedra. We shall show that this and only this assignment 
yields spherical functions at the triangular faces of the relativistic spin network. Geometric sharing of simplicial boundaries is
transcribed into the intertwining condition of the fields.
For two tetrahedra sharing a triangular face it follows that the
corresponding irreps must intertwine, i.e. can be coupled to the
identity irrep of $SL(2,C)$. This coupling was analyzed in general in the
previous subsection. We expand the labels in eq. \ref{e43}
according to 
\begin{equation}
\label{e49}
\lambda \mu \rightarrow \lambda \mu \downarrow(00),\; 
\lambda^C \mu^C 
\rightarrow \lambda^C \mu^C \downarrow(00),
\end{equation}
and in addition drop the left-hand  multiplicity label $\mu$ since for the irreps of the groups $G=(SO(4,R), SL(2,C)$, the 
identity irrep $J=0$ of the stability group $SU(2)$ occurs once and only once.

\medskip
\noindent
{\bf 15 Def}: Zonal spherical functions for the irrep $\lambda$ of 
$G= (SO(4,R); SL(2,C))$ with subgroup $SU(2)$, in a basis where $G$  is explicitly reduced 
with respect to this subgroup, are  defined by
\begin{equation}
\label{e50}
f^{\lambda}(g):=
\langle \lambda  \downarrow(00)| T^{\lambda}_{g}|
\lambda  \downarrow(00)\rangle.
\end{equation}
Clearly the spherical functions eq. \ref{e50} are invariant under the substitution
$g \rightarrow u_1gu_2,\; (u_1,u_2) \in SU(2)$.
Therefore the zonal spherical functions eq. \ref{e50} can be assigned 
to the double cosets $SU(2) \backslash SL(2,C) /SU(2)$ and must be
distinguished from the spherical harmonics eq. \ref{e48}. The spherical functions
eq. \ref{e50} are the relativistic counterparts  of the ones considered in part I section 5.2.

Applying now eq. \ref{e43} we find for the intertwining of
two spherical harmonics an expression in terms of a zonal spherical 
function eq. \ref{e50},
\begin{eqnarray}
\label{e51}
&& \left[
\sum_{\nu}
\langle \lambda  \downarrow(00)| T^{\lambda}_{g_1}|\lambda \nu\rangle
\langle \lambda^C  \downarrow(00) 
| T^{\lambda^C}_{\sigma_2(g_2)}|\lambda^C \nu^C\rangle
\right]
\\ \nonumber
&&= f^{\lambda}(g_1g_2^{-1})=
\langle \lambda  \downarrow(00)| T^{\lambda}_{g_1g_2^{-1}}
|\lambda  \downarrow(00)\rangle.
\end{eqnarray}
This expression is invariant under the right action\\
 $(g_1,g_2) \rightarrow (g_1q, g_2q),\;
q \in SL(2,C)$.

The Wigner coefficients have been suppressed in view of eq. \ref{e47}.
The expression in the last line defines  a zonal spherical function with respect 
to $SL(2,C)>SU(2)$. 

\medskip
\noindent
{\bf 16 Prop}: The interwining of  two spherical harmonics on $SL(2,C)/SU(2)$
belonging to conjugate irreps $\lambda,\lambda^C$   yields a single 
zonal spherical  function eq. \ref{e51} of the product $g_1g_2^{-1}$ of group elements.

We shall derive the spherical functions for $SL(2,C)>SU(2)$ in section 8.\\
We add four remarks to this result:\\
{\bf  1: Conjugate irreps}. Although the two intertwining irreps are $\lambda, \lambda^C$, 
the spherical function in eq. \ref{e51}
belongs to the single irreps $\lambda$. This asymmetry is removed upon 
considering the inverse element $g_2g_1^{-1}$ which by eq. \ref{e42} 
leads to a spherical function belonging to the irrep $\lambda^C$.\\
{\bf 2: Reduction to $SU(2)$}. The occurrence of a single irrep $\downarrow(00)$ of $SU(2)$
puts a constraint on the irrep $\left[ m, \rho \right]$, [2] p. 567.
In particular the irreps $\left[m,0\right]$ are ruled out, see [4] p. 163.\\
{\bf 3: Self-conjugacy for irreps of $SL(2,C)$}. In the application to quantum gravity,
the intertwining condition at a 
triangular face of a simplex in general requires two conjugate and hence inequivalent irreps. Since a simplex in $M(1,3)$ is bounded by five
tetrahedra, the pairwise intertwining of irreps for these tetrahedra 
implies a single self-conjugate irrep at each tetrahedron of a simplex.
 
In two special cases, $\lambda =\left[ m, 0\right]$ or 
$\lambda =\left[ 0, \rho \right]$, the irrep and its conjugate are 
equivalent. This follows from eq. \ref{e47} and from the 
equivalence $\left[m, \rho\right] \sim \left[-m, -\rho\right]$ stated
with proof in [4] p. 168, in accord with identical eigenvalues eq. \ref{e14} of the
two Casimir operators. The irrep $\left[m,0\right]$ is ruled out as explained in 
{\bf 2}. The restriction to the special irreps $\left[0,\rho\right],\: \left[0,\sigma\right]$ of $SL(2,C)$
would be the relativistic analog  of  the notion of simple irreps as used in quantum gravity, see section 5 of part I  [1]. In [8] p. 3105 the authors argue that 
the complementary series irreps have Plancherel measure zero and therefore 
can be neglected for relativistic spin networks.

{\bf 4: Self-conjugacy for irreps of $SO(4,R)$}. In case of $SO(4,R)$ considered in part I [1],
the irreps can be expressed by pairs of irreps  of $SU(2)$. Therefore
the conjugate irreps obey $\overline{D}^{(j_1,j_2)}\sim D^{(j_1,j_2)}$
and so are self-conjugate. 
This applies in particular to simple  irreps $D^{(j_2j_2)}$.

\section{Zonal spherical function for the group SO(1,3,R).}

Given the unitary representation $T_g$ of the group SL(2,C) and the identity representation $(JM)=(00)$ of the subgroup SU(2), the zonal spherical function is defined
as in Definition 15 eq. \ref{e50}.

From the properties of the spherical function it is sufficient to take the representation of the Lorentz boost. From eq. \ref{e59} we have with $M=M'= 0, J=J'=0$
\begin{eqnarray}
\label{e60}
&&f^{[m,\rho]} (\tau )=\left\langle {\psi_{JM=00} |  \exp (- iK_3 \tau)  |
\psi_{J'M'=00}} \right\rangle
\\ \nonumber
&&=\int {d_0^{-2}}e^{-i\lambda \tau }\left[
{p_0^{(0,0)}} \right]^2\omega (\lambda )d\lambda 
\\ \nonumber
&&=\int\limits_{-\infty }^\infty  {{{e^{-i\lambda \tau }} \over {\left| {\Gamma (1+i\rho )} \right|^2}}}{1 \over {4\pi
}}\left| {\Gamma \left( {{1 \over 2}+i{{\lambda +\rho } \over 2}} \right)\Gamma \left( {{1 \over 2}+i{{\lambda -\rho }
\over 2}} \right)} \right|^2d\lambda 
\\ \nonumber
&&=\int\limits_{-\infty }^\infty  {e^{-i\lambda \tau }}{{\sinh(\pi \rho) } \over {4\rho }}{1
\over {\cosh(\pi \left( {{{\lambda +\rho } \over 2}} \right))}}
{1 \over {\cosh(\pi \left( {{{\lambda -\rho } \over 2}}
\right))}}d\lambda
\end{eqnarray}

\noindent where we have used the properties of ${\Gamma }$ functions. From the residue theorem at the poles $\lambda
=\mp \rho +\left( {2n+1} \right)i,\;\;n=0,1,2,\dots,$ we easily obtain for $\tau<0$

\begin{equation}
\label{e61}
f^{[0,\rho]}(\tau )=\: i\;\frac{\exp(i\rho \tau)-\exp(-i\rho \tau )}{\rho }\:\exp \tau\: \sum\limits_{n=0}^\infty  (\exp(2\tau))^n =\frac{1}{\rho}
\frac{\sin(\rho \tau)}{\sinh(\tau)}
\end{equation}
for the principal series with $l_0=0,\;l_1=i\rho $, see  [4] p. 166.
Identical expressions can be obtained for $\tau>0$ if we apply the residue theorem to eq. \ref{e60} at the poles $\lambda=\mp \rho -(2n+1)i,\; n=0,1,2,...$  In the  
application of the residue theorem, we have to check that the integrand of 
eq. \ref{e60} goes to zero when $|\lambda| \rightarrow \infty$ in the upper half-plane
for $\tau<0$, or in the lower half-plane for $\tau>0$. This can be proven
very easily in general and in the particular case of $\lambda= \mp \rho +(2n+1)i,
n=0,1,2,...$ by  L' Hospital 's rule. 

We obtain in the same way,
see [4] p. 186,
\begin{equation}
\label{e62}
f^{[0,\sigma]}(\tau )={1 \over \sigma }{{\sinh(\sigma \tau) } \over {\sinh(\tau) }}
\end{equation}
for the complementary series, with $l_0=0,\;l_1=\sigma ,\;\left| \sigma  \right|<1$.

\section{A SO(1,3,R) invariant for the state sum of a spin foam model} 

As in the case of euclidean $SO(4)$ invariant model, we take a non degenerate finite triangulation of a 4-manifold. We consider
the 4-simplices in the homogeneous space $SO(1,3,R)/SO(3)$ $\backsim H^3$, the hyperboloid $\left\{ {\left. x \right|x\cdot
x=1,x^0>0} \right\}$ and define the bivectors $b$ on each face of the 4-simplex, that can be space-like, null or timelike
($\left\langle {b,b} \right\rangle >0,=0,<0$, respectively).

In order to quantize the bivectors, we take the isomorphism $b=*L\left( {b^{ab}={1 \over
2}\varepsilon^{abcd}L_d^eg_{ec}} \right)$ with $g$ a Minkowski metric.

The condition for $b$ to be a simple bivector $\left\langle {b,*b} \right\rangle =0$, gives $C_2=\left\langle {L,*L}
\right\rangle =\vec J\cdot \vec K=m \rho =0$

We have two cases:

1) $\rho =0,\quad C_1=\left\langle {L,L} \right\rangle =\vec J^2- \vec K^2=m ^2-1>0$; $L$, space-like, $b$ time-like,

2) $m =0,\quad C_1=\vec J^2- \vec K^2=-\rho ^2-1<0$; $L$, time like, $b$ space like (remember, the Hodge operator $*$ changes the
signature)

In case 2) $b$ is space-like, $\left\langle {b,b} \right\rangle >0$. Expanding this expression in terms of space-like vectors
$x,y,$ 
\begin{eqnarray}
\label{e63}
b_{\mu \nu }b^{\mu \nu }&=&\left( {x_\mu y_\nu -x_\nu y_\mu } \right)\left( {x^\mu y^\nu -x^\nu y^\mu } \right)=\\
\nonumber &=&\left\| x
\right\|^2\left\| y \right\|^2-\left\| x \right\|^2\left\| y \right\|^2\cos ^2\eta \left( {x,y} \right)=\left\| x
\right\|^2\left\| y \right\|^2\sin ^2\eta \left( {x,y} \right)
\end{eqnarray}
where $\eta \left( {x,y} \right)$ is the Lorentzian space-like distance between $x$ and $y$; this result gives a geometric
interpretation between the parameter $\rho$ and the area expanded by the bivector $b=x\wedge y$, namely, $\left\langle {b,b}
\right\rangle =\mbox {(area)}^2\left\{ {x,y} \right\}=\left\langle {*L,*L} \right\rangle \cong \rho ^2$. This result is the
analogue to that obtained in Section 7 of part I [1] where the area of the triangle expanded by the bivector was proportional to the value $(2j+1)$,
$j$ being the spin of the representation.

In order to construct the Lorentz invariant state sum we take a non-degenerate finite triangulation in a 4-dimensional simplices in such a way that all
3-dimensional and 2-dimensional subsimplices have space-like edge vectors which span space-like subspaces (Prop 7). We attach to each 2-dimensional face a simple
irrep. of SO(1,3,R) characterized by the parameter $[0, \rho]$. For these simple representations the intertwining condition is preserved (Prop 16 and remark  {\bf 3}).

The state sum is given by the expression
\begin{equation}
\label{e64}
Z = \int\limits_{\rho  = 0}^\infty d\rho \prod_{\rm triangle} \rho^2 \prod_{\rm tetra} {\Theta _4 } \left( {\rho '_1 , \cdots ,\rho '_4 } \right)\prod\limits_{\rm 4 - simplex}
{I_{10}
\left( {\rho ''_1 , \cdots ,\rho ''_{10} } \right)} 
\end{equation}
where $\rho$ refers to all the faces in the triangulation, ${\rho '}$ corresponds to the simple irreps attached to 4 triangles in the tetrahedra and ${\rho ''}$ 
corresponds to the simple irrep attached to the 10 triangles in the 4-simplices.

The functions $\Theta _4$ and $I_{10}$ are defined as traces of recombination diagrams for the simple representations. The expressions eq. \ref{e64} are explicitely given as multiple integrals over the upper sheet $H^3$ of
the 2-sheeted hyperboloid in Minkowski space. For the integrand we take the zonal 
spherical functions eqs. \ref{e61}, \ref{e62},
\begin{equation}
\label{e65}
f^{[0,\rho]}  \left( {x,y} \right) = {1 \over \rho }\;{{\sin( \rho\, \tau \left( {x,y} \right))} \over {\sinh (\tau \left( {x,y} \right))}},\:
f^{[0,\sigma]}  \left( {x,y} \right) = {1 \over \sigma }\;{{\sinh( \sigma\, \tau \left( {x,y} \right))} \over {\sinh (\tau \left( {x,y} \right))}}
\end{equation}
for the  chosen irrep of $SL(2,C)$ where  $\tau (x,y)$ is the hyperbolic distance between $x$ and $y$. If in remark {\bf 3} after Prop. 16 we follow  [8] p. 3105 we can drop the second expression of eq. \ref{e65} for the supplementary series. \\

The contribution  of a recombination diagram is given by a multiple integral of products of spherical functions. 

For a tetrahedron we have
\begin{equation}
\label{e66}
\Theta _4 \left( {\rho '_1 , \cdots ,\rho '_4 } \right) = {1 \over {2\pi ^2 }}\int\limits_H {f^{\rho'_1 } } \left( {x,y} \right) \cdots f^{\rho '_4 } \left( {x,y} \right)dy
\end{equation}
where we have dropped one integral for the sake of normalization without loosing Lorentz symmetry.

For a 4-simplex we have
\begin{equation}
\label{e67}
I_{10} \left( {\rho _1 , \cdots ,\rho _{10} } \right) = {1 \over {2\pi ^2 }}\int\limits_{H^4 } {\prod\limits_{i < j}^5 {f^{\rho _{ij} } } } \left( {x_i,
x_j } \right)dx_1 dx_2 dx_3 dx_4 
\end{equation}
Equations \ref{e65} to \ref{e67} define the state sum completely, that has been proved to be finite [9].

The asymptotic properties of the spherical functions are related to the Einstein-Hilbert action [1] giving a connection of the model with the theory of general relativity.

\acknowledgements
One of the author (M.L.) expresses his gratitude to the Director of the Institut f\"{u}r Theoretische Physik, of the University of T\"ubingen, where part of
this work was done with financial support from M.E.C. (Spain) with Grant: FPA2006-09199.

\end{article}

\begin{thebibliography}{99}
\bibitem[\protect\citeauthoryear{Lorente and Kramer}{}]{LorenteAndKramer}
[1] Lorente, M. and P. Kramer. ``Discrete Quantum Gravity: I. Spherical functions of the representation of the SO(4) group with respect 
to the SU(2) subgroup and its application to the Euclidean invariant weight 
for the Barrett-Crane model''.
\end{thebibliography}
\end{document}